\documentclass[aps,twocolumn,pra]{revtex4}
\usepackage{epsfig,rotating}
\usepackage{amsmath,amssymb,amsthm}
\usepackage{graphicx}
\usepackage{psfrag}
\usepackage{bm}
\usepackage[dvips]{color}

\begin{document}

\title{Husimi distribution, Wehrl entropy and superradiant phase in spin-boson interactions}

\author{R. del Real}
\affiliation{Departamento de F\'{\i}sica At\'omica, Molecular y Nuclear, Universidad de Granada, Fuentenueva s/n, 18071 Granada,
Spain}
\author{M. Calixto}
\affiliation{Departamento de Matem\'atica Aplicada, Universidad de Granada,
Fuentenueva s/n, 18071 Granada, Spain}
\author{E. Romera}
\affiliation{Departamento de F\'{\i}sica At\'omica, Molecular y Nuclear and
Instituto Carlos I de F{\'\i}sica Te\'orica y
Computacional, Universidad de Granada, Fuentenueva s/n, 18071 Granada,
Spain}

\date{\today}
\begin{abstract}
We study the Husimi distribution of the ground state in the Dicke model of field-matter interactions to visualize the
quantum phase transition, from normal to superradiant, in phase-space. We follow  an exact 
numerical  and  variational analysis,  without making use of the usual 
Holstein-Primakoff approximation. We find 
that Wehrl entropy of
the Husimi distribution provides an indicator of  the sharp change of 
symmetry
trough the critical point. Additionally, we note that the zeros of the 
Husimi distribution characterize the Dicke model quantum phase 
transition.

\end{abstract}
\maketitle

\section{Introduction}
The study of quantum phase transitions (QPTs) is an important subject in
many-body quantum physics \cite{sachdev}. If we consider a quantum system
described by the Hamiltonian $H=H_0+\lambda H_1$, where $H_0$ and $H_1$ 
have different symmetries and $\lambda$ is a control parameter, QPT occurs when
$\lambda$ reaches a critical value $\lambda_c$ for which the properties 
of the system change suddenly.

In this work we will analyze phase-space properties for a QPT and,  for this
purpose, we will consider the representative Dicke model of spin-boson
interactions (see e.g. \cite{emaryprl2,emarypra,emarypre,brandes}). There are 
several distributions to analyze
phase-space properties \cite{Gerry2005}; the most popular one  is the Wigner
distribution, but there is another important one, the Husimi
distribution, which has the interesting property of non-negativity and it is
defined  as the overlap between a minimal uncertainty (coherent) state and
the wavefunction. Recently, we have proposed the Husimi distribution as a 
tool for a phase-space visualization of QPTs using two  algebraic models to exemplify the study: the Dicke  model \cite{husidi} and the  
vibron model \cite{husivi}, this last used to study rotational and vibrational spectra in
diatomic and polyatomic molecules, which also exhibit a (shape) QPT. In Ref. \cite{husidi} we made use of 
the Holstein-Primakoff approximation \cite{HP} (large spin $j$) to approximate the atomic 
sector by an harmonic oscillator for a large number of atoms $N=2j$. Here we won't use this approximation and work with finite $N$ in an exact manner. 

   The advantage of working in phase space is that
we can analyze contributions in position and momentum space jointly.
  Additionally, we have characterized QPTs using
  the zeros of the Husimi distribution. Other information theoretic measures for 
  QPT's in the Dicke
and vibron models have been recently studied  in position
and/or momentum spaces, separately. In particular, it has been shown that
there is an abrupt change of the R\'enyi entropy \cite{pla11}, Fisher
information \cite{physica12} and complexity measures \cite{jme11} at the
transition point in the Dicke model. Moreover, it has been found that
uncertainty Shannon \cite{epl2012} and R\'enyi \cite{renyipra,renyinagys} entropic
relations  accounts for the QPTs better than other variance-based uncertainty relations. See also \cite{jpa12}
 for a recent paper on vibration-rotation entanglement measures of vibron models in the `rigidly bent' phase.

The structure of the paper is the following. In Section \ref{sec1}  we 
briefly remind the Dicke model, boson and spin-$j$ coherent states 
 and we  present the Husimi distribution (without the
Holstein-Primakoff approximation) and the Wehrl entropy. In
Section \ref{sec2} we will present numerical and variational results in terms of symmetry-adapted 
coherent states. Three-dimensional plots, contour lines and Wehrl entropy of the Husimi distribution reveal a drastic change in the symmetry 
of the ground state wave function and provide a signature for the QPT even for a finite number of particles.  
Finally, zeros of the Husimi  distribution (in the 
variational approximation) are also
computed and graphically represented  to characterize the QPT.

\section{Dicke Hamiltonian, Husimi distribution and Wehrl entropy}\label{sec1}

The single-mode  Dicke model is a well studied object in the field of QPTs \cite{emaryprl2,emarypra,brandes}. 
In this case the Hamiltonian is given by 
\begin{equation}
\label{qpt01}
H=\omega_0 J_z + \omega a^{\dag} a + \frac{\lambda}{\sqrt{2 j}}
( a^{\dag} + a )( J_+ + J_-) ,
\end{equation}
describing an ensemble of $N$ two-level atoms with level-splitting $\omega_0$, with 
$J_z$, $J_{\pm}$ the  angular momentum operators for a
pseudospin of length $j=N/2$, and $a$ and $a^{\dag}$ are the 
bosonic operators of
the field with frequency $\omega$. It is well known that
there is a QPT at the critical value of the coupling parameter
$\lambda=\lambda_c=\frac{\sqrt{\omega\omega_0}}{2}$ from 
the so-called normal phase ($\lambda<\lambda_c$)  to the superradiant phase
($\lambda > \lambda_c$). 
  
Let us consider a basis set $\left\{|n;j,m\rangle\equiv |n\rangle\otimes|j,m\rangle \right\}$ of
the Hilbert space, with $\left\{|n\rangle\right\}_{n=0}^{\infty}$ the number
states of the field and $\left\{|j,m\rangle\right\}_{m=-j}^{j}$ the so called
Dicke states of the atomic sector. The matrix elements of the Hamiltonian in this basis are:
\begin{eqnarray}
\langle n^{\prime};
j^{\prime},m^{\prime}|H|n;j,m\rangle= (n\omega+m\omega_0)
\delta_{n^{\prime},n}\delta_{m^{\prime},m}
\notag\\ 
+\frac{\lambda}{\sqrt{2j}}(\sqrt{n+1}\delta_{n^{\prime},n+1}+\sqrt{n}\delta_{n^{\prime},n-1})\notag \\ \times (\sqrt{j(j+1)-m(m+1)}
\delta_{m^{\prime},m+1}\notag\\ +\sqrt{j(j+1)-m(m-1)}\delta_{m^{\prime},m-1}).\label{hamel}
\end{eqnarray}
At this point it is important to note that time evolution preserves the parity $e^{i\pi(n+m+j)}$ of a given state 
$|n;j,m\rangle$. That is, the parity operator $\hat\Pi=e^{i\pi(a^\dag a+J_z+j)}$ commutes with ${H}$ and both operators 
can then be jointly diagonalized. In particular, the ground state must be even (see later on Eq. (\ref{sacs})).

Let us denote by 
\begin{equation}
\begin{array}{l}
|\alpha\rangle=e^{-|\alpha|^2/2}e^{\alpha a^\dag}|0\rangle=
e^{-|\alpha|^2/2}\sum_{n=0}^\infty\frac{\alpha^n}{\sqrt{n!}}|n\rangle,\\
|z\rangle=(1+|z|^2)^{-j}e^{zJ_+}|j,-j\rangle=\\(1+|z|^2)^{-j}\sum_{m=-j}^j\binom{2j}{j+m}^{1/2}z^{j+m}|j,m\rangle,
\end{array}\label{cohs}
\end{equation}
(with $\alpha,z\in\mathbb C$) the  standard (canonical or Glauber) and spin-$j$ Coherent States (CSs) for the photon and the
particle sectors, respectively. It is well known (see e.g. \cite{Perelomov}) that  coherent states form an overcomplete set 
of the corresponding Hilbert space and fulfill 
the closure relations or resolutions of the identity:
\begin{eqnarray}
1&=&\frac{1}{\pi}\int_{{\mathbb R}^2}|\alpha\rangle\langle\alpha|d^2\alpha, \nonumber\\ 
1&=&\frac{2j+1}{\pi}\int_{{\mathbb R}^2}|z\rangle\langle z|\frac{d^2 z}{(1+|z|^2)^2},\label{closure}
\end{eqnarray}
with $d^2w\equiv d\mathrm{Re}(w)d\mathrm{Im}(w)$ (or $d^2w=rdrd\theta$ in polar coordinates $w=r e^{i\theta}$) the Lebesgue measure on $\mathbb C$. 
The complex parameters $\alpha$ and $z$ are related to the mean number of photons, as $\langle\alpha|a^{\dag} a|\alpha\rangle=|\alpha|^2$,  
and the mean fraction of excited atoms, as $\langle z|J_z+j|z\rangle=N|z|^2/(1+|z|^2)$, respectively.  
It is also  
straightforward to see that the probability amplitude of detecting $n$ photons and $j+m$ excited atoms in 
$|\alpha,z\rangle\equiv |\alpha\rangle\otimes|z\rangle$ is given by:
\begin{equation}
 \varphi_{n,m}^{(j)}(\alpha,z)=\langle n|\alpha\rangle\langle j,m|z\rangle=\frac{e^{-|\alpha|^2/2}\alpha^n}{\sqrt{n!}}
\frac{\sqrt{\binom{2j}{j+m}}z^{j+m}}{(1+|z|^2)^j}.\label{probamp}
\end{equation}
The ground state vector $\psi$ will be given as an expansion 
\begin{equation}
 |\psi\rangle=\sum_{n=0}^{n_c}\sum_{m=-j}^{j} c_{nm}^{(j)}|n;j,m\rangle
\end{equation}
where the coefficients $c_{nm}^{(j)}$ are calculated by numerical diagonalization of \eqref{hamel}, with a given cutoff $n_c$, and depend on the control 
parameter $\lambda$. The 
Husimi distribution of $\psi$ is then given by
\begin{eqnarray}
 {\Psi}(\alpha,z)&=&|\langle\alpha,z|\psi\rangle|^2\label{husiz}\\ &=&\sum_{n,n'=0}^{n_c}\sum_{m,m'=-j}^{j} c_{nm}^{(j)} 
\bar c_{n'm'}^{(j)}\varphi_{n,m}^{(j)}(\alpha,z)\varphi_{n',m'}^{(j)}(\bar\alpha,\bar z)\nonumber
\end{eqnarray}
and normalized according to:
\begin{equation}
 \int_{\mathbb R^4} {\Psi}(\alpha,z) d\mu(\alpha,z)=1, 
\end{equation}
with integration measure: 
\begin{equation}
 d\mu(\alpha,z)=\frac{2j+1}{\pi^2}\frac{{d^2\alpha}\,d^2 z}{(1+|z|^2)^2}.
\end{equation}
An important quantity to visualize the QPT in the Dicke model across the critical point $\lambda_c$ will be the 
Wehrl entropy
\begin{equation}
W_{j}(\lambda)=-\int_{{\mathbb R}^4} \Psi(\alpha,z)\ln(\Psi(\alpha,z))\; d\mu(\alpha,z),\label{wehrl1}
\end{equation}
where the dependence of $W_{j}$ on $\lambda$ comes from the dependence of $c_{nm}^{(j)}$ on $\lambda$.

\section{Numerical  versus  variational results}\label{sec2}

\begin{figure}
\includegraphics[width=4cm]{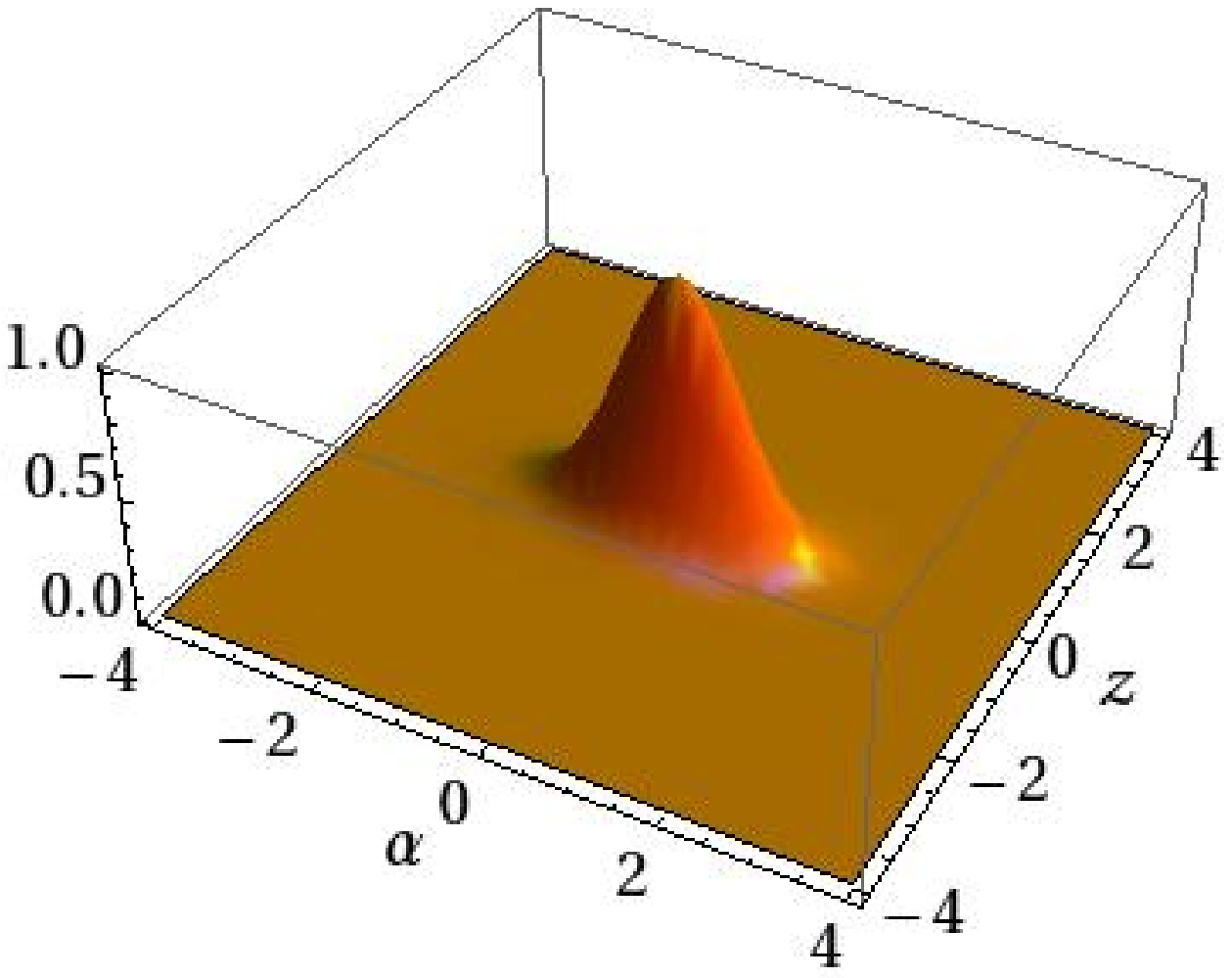}
\includegraphics[width=4cm]{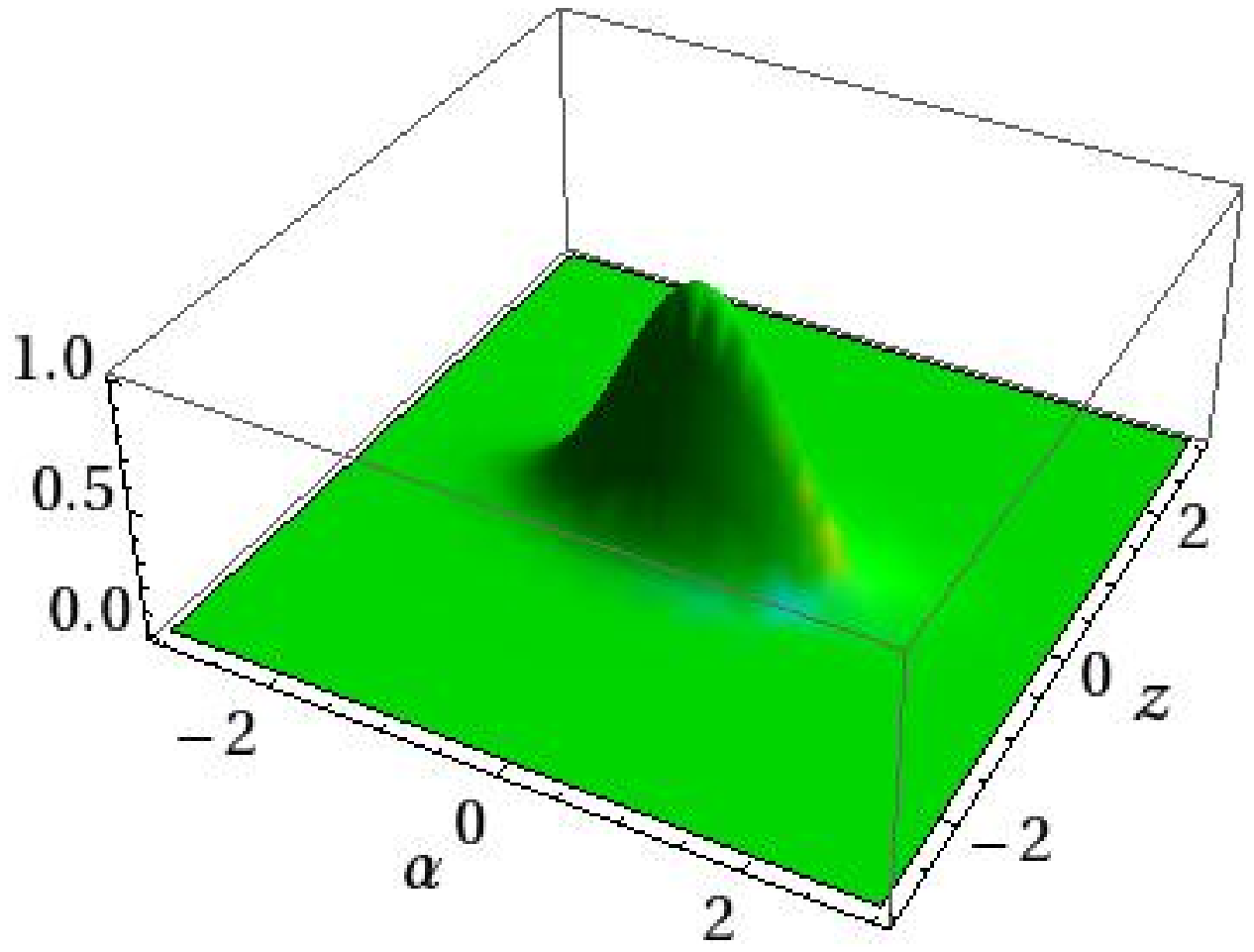}
\includegraphics[width=4cm]{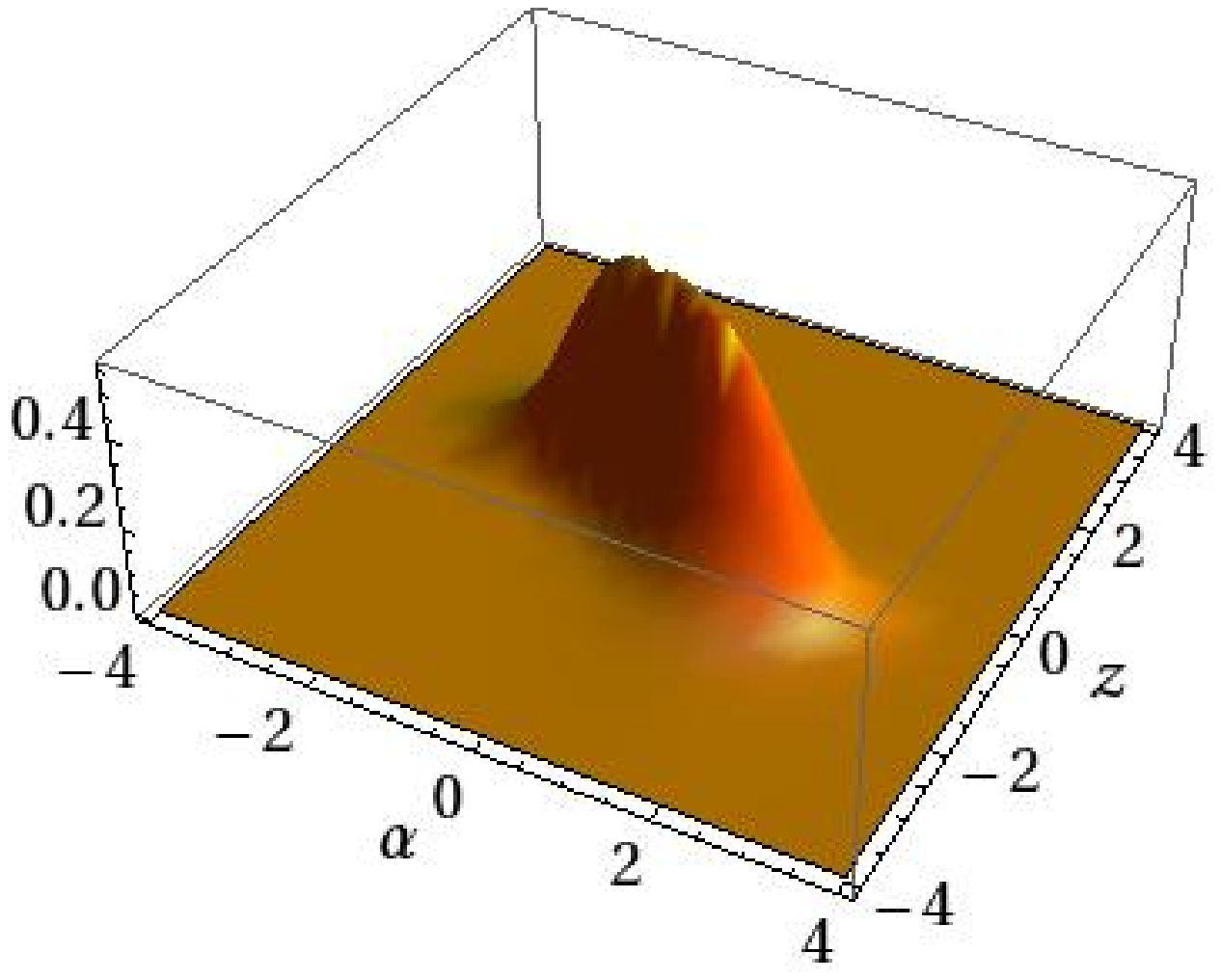}
\includegraphics[width=4cm]{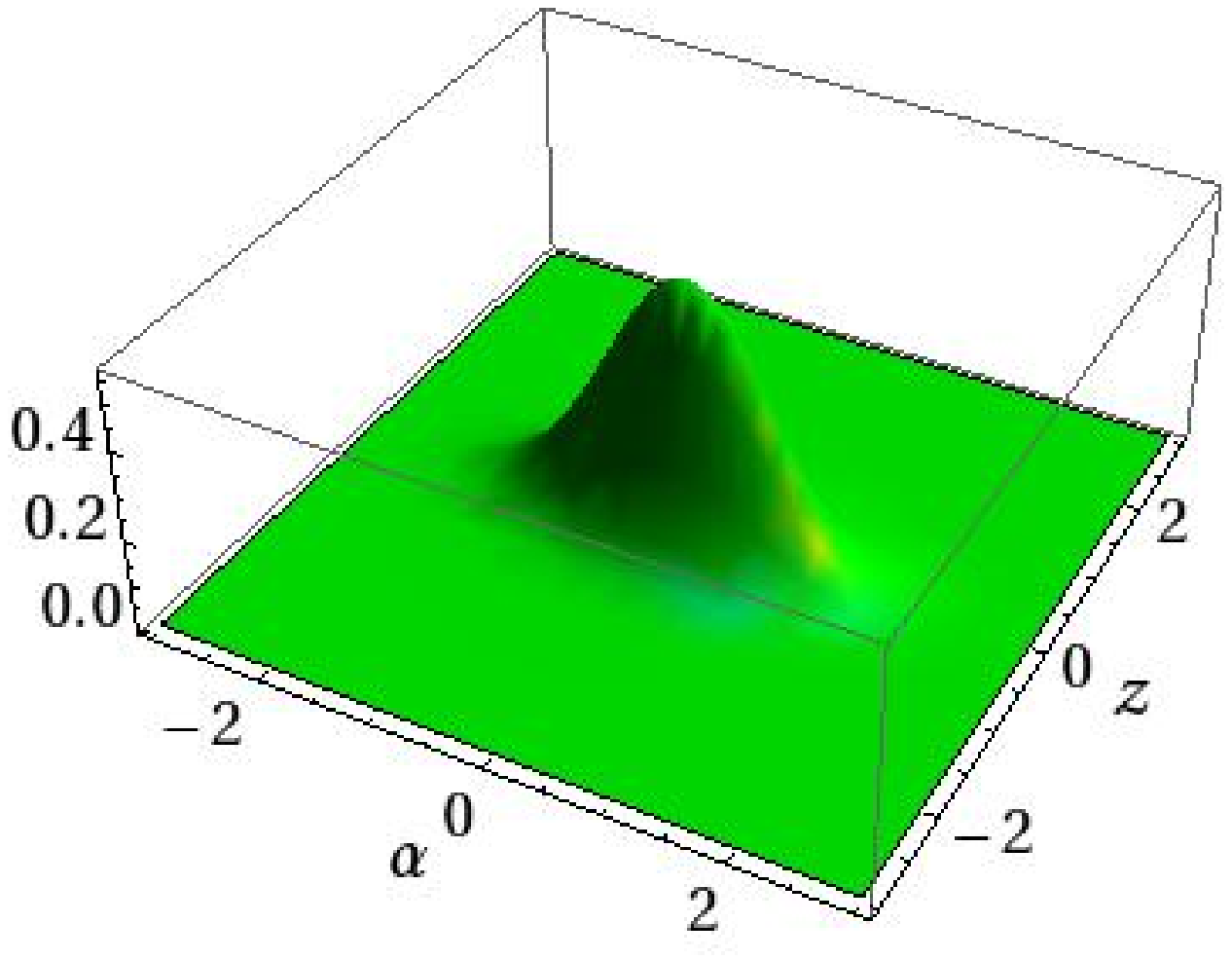}
\includegraphics[width=4cm]{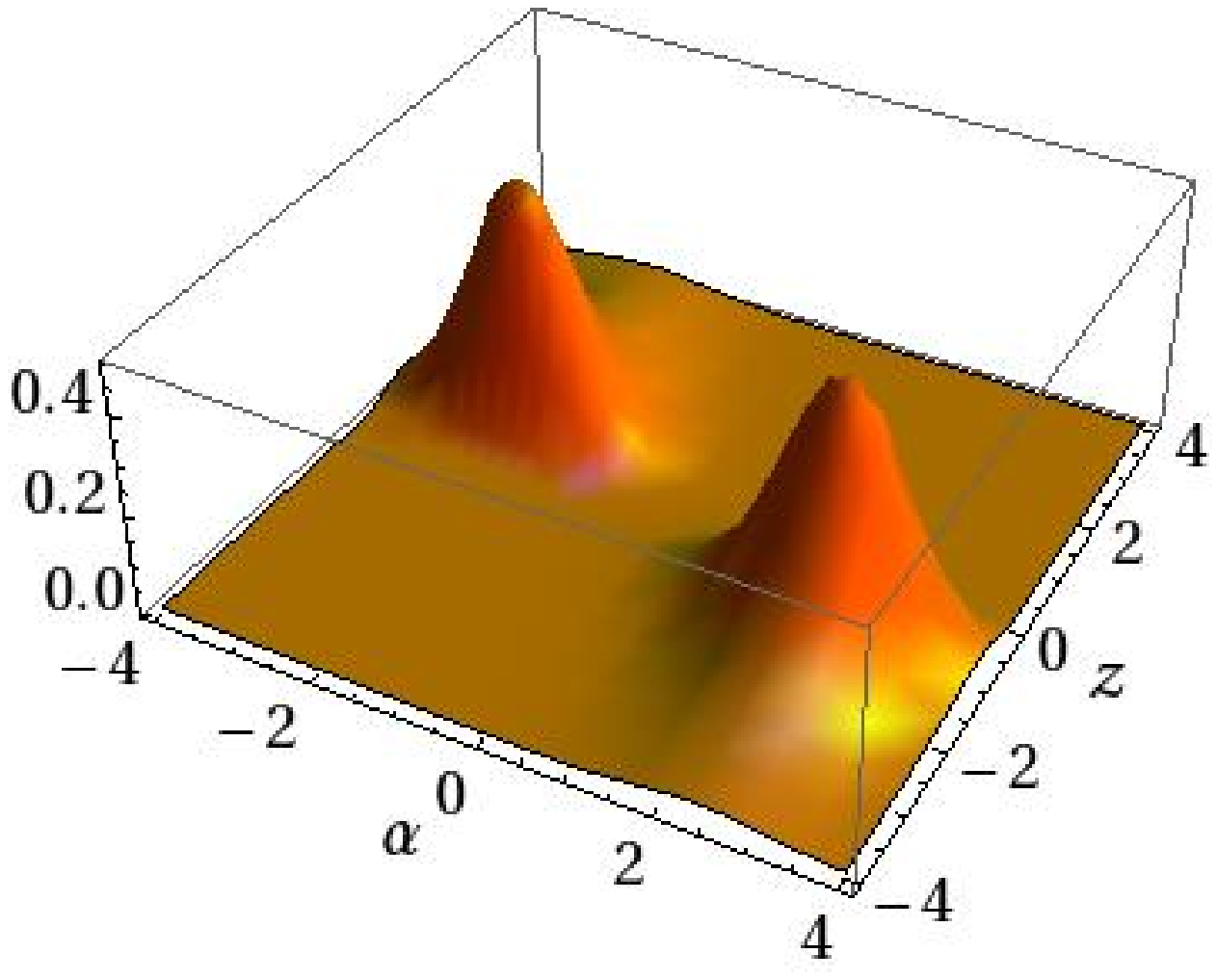}
\includegraphics[width=4cm]{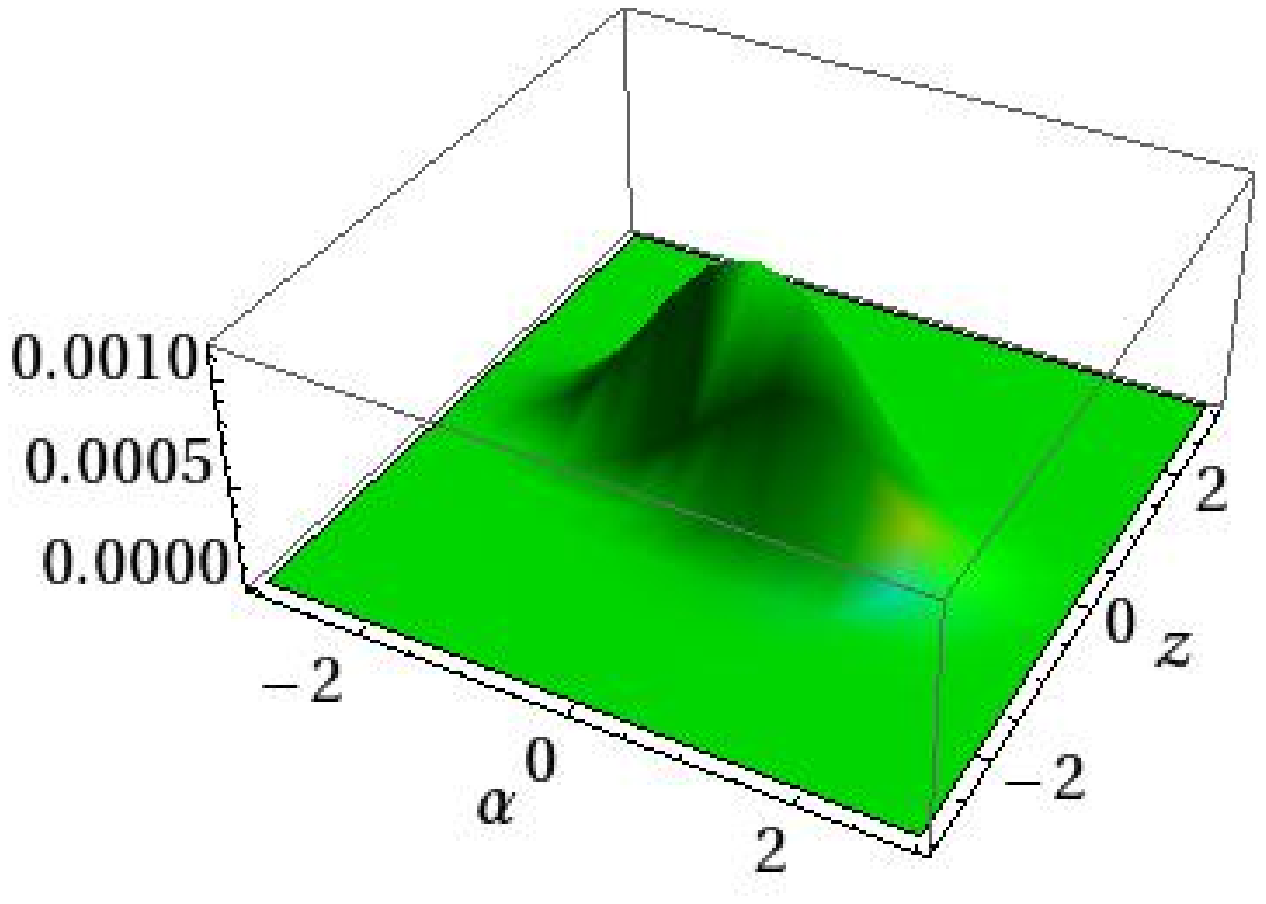}
\caption{(Color online) 3D-Plot   of the exact Husimi distribution in
  (left) ``position space''  ($\alpha$ and $z$ real), and (right) ``momentum space''  ($\alpha$ and $z$ imaginary)
for different values of $\lambda$ (from
top to bottom: $\lambda=0$, $\lambda=0.6$ and $\lambda=1$) for $j=3$ and $\omega=\omega_0=1\Rightarrow \lambda_c=0.5$. Atomic units.}
\label{husimifig}
\end{figure}

\begin{figure}
\includegraphics[width=4cm]{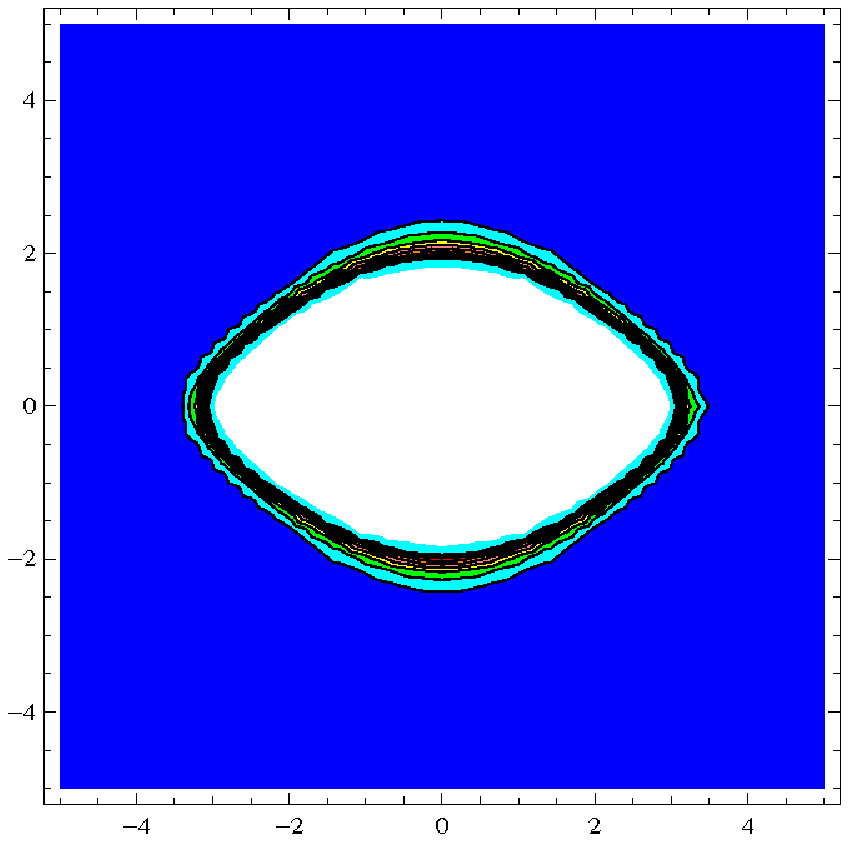}
\includegraphics[width=4cm]{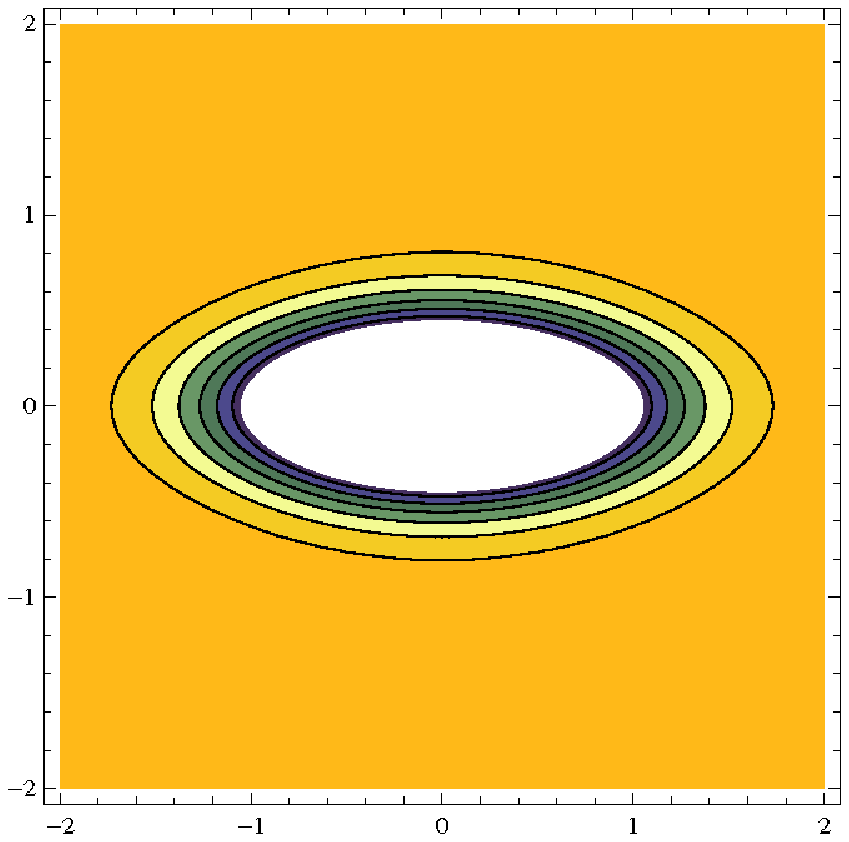}
\includegraphics[width=4cm]{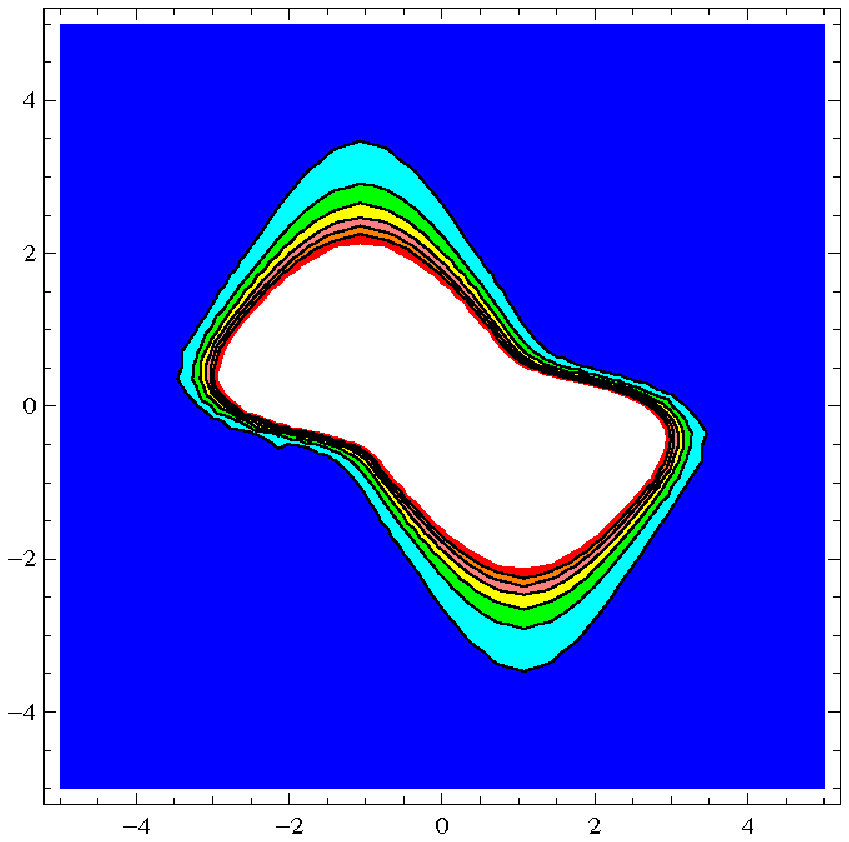}
\includegraphics[width=4cm]{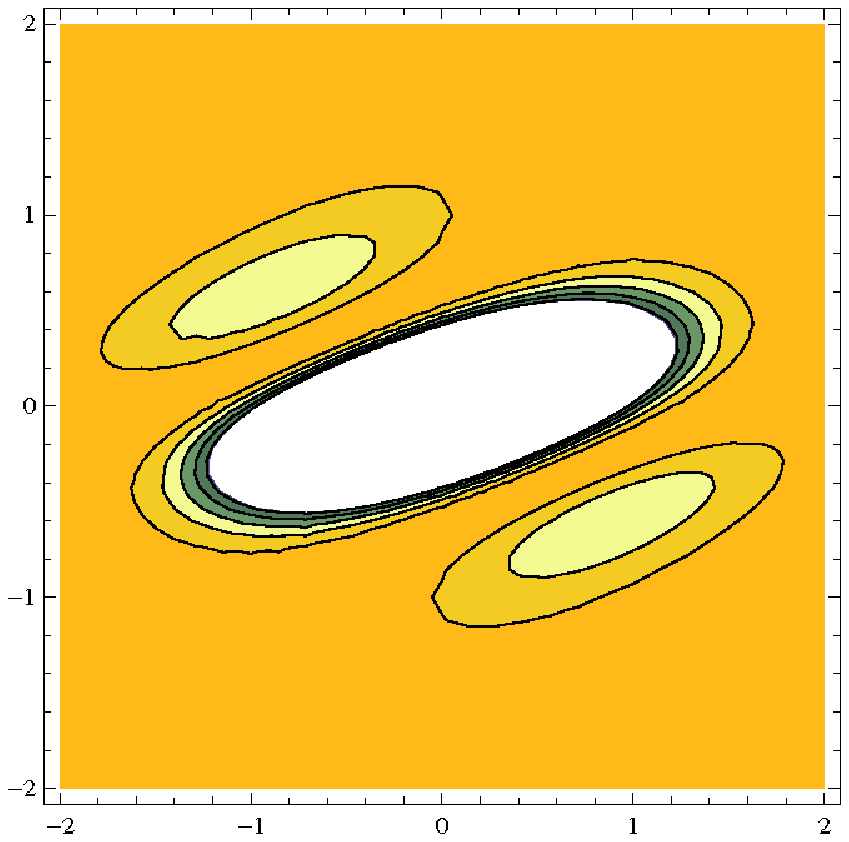}
\includegraphics[width=4cm]{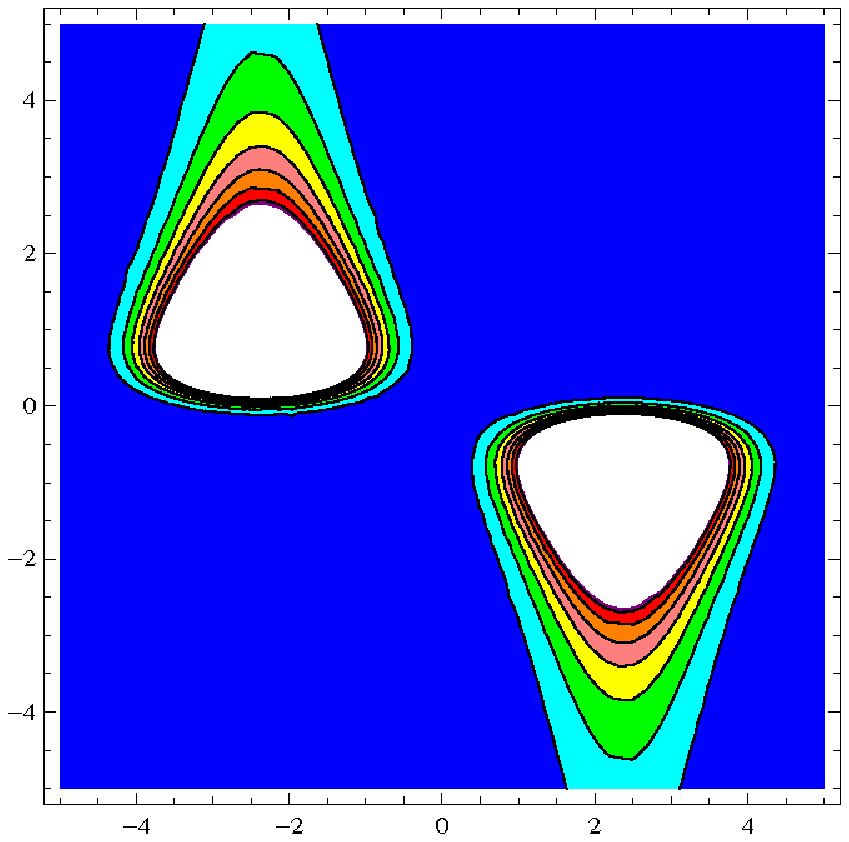}
\includegraphics[width=4cm]{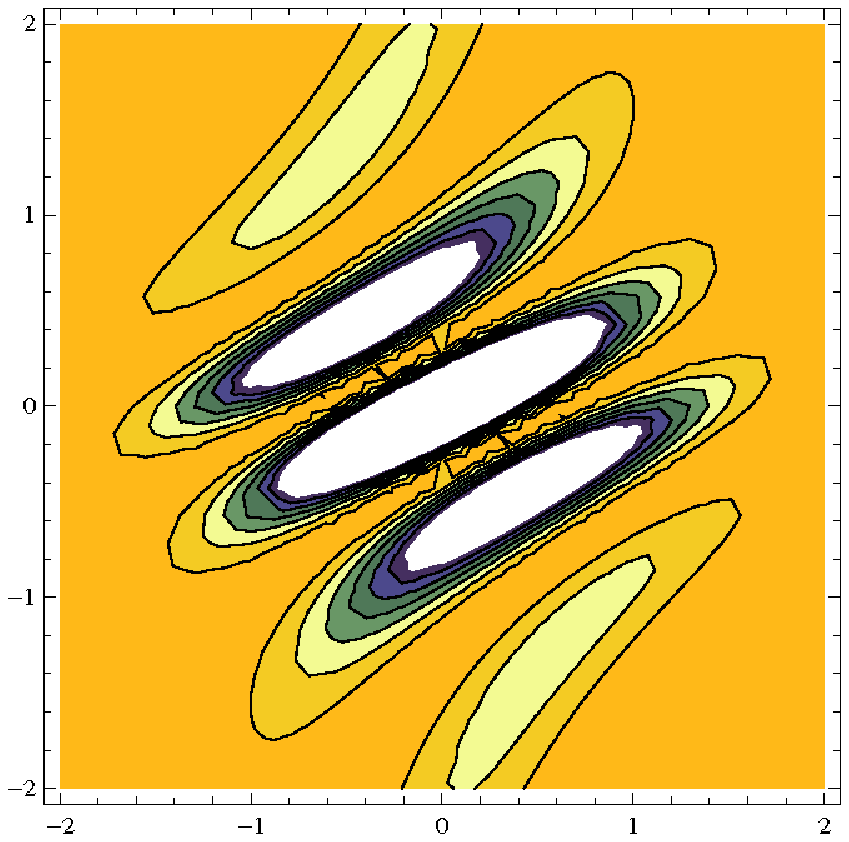}
\caption{(Color online)  Contour lines of the variational Husimi distribution $\Psi_+(\alpha,z)$ in ``position space'' ($\alpha$ and $z$ real; left panel) and 
``momentum space'' ($\alpha$ and $z$ imaginary; right panel) 
  for different values of $\lambda$ (from
top to bottom: $\lambda=0$, $\lambda=0.6$ and $\lambda=1$) for $j=3$ and $\omega=\omega_0=1\Rightarrow \lambda_c=0.5$. Atomic units.}
\label{husimifig2}
\end{figure}

In Figure \ref{husimifig} we represent a 3D plot of the exact Husimi distribution of the ground state $\Psi(\alpha,z)$  in `position' ($\alpha$ and $z$ real) and  
`momentum' ($\alpha$ and $z$ imaginary) spaces. We observe that the Husimi distribution in position space is concentrated around $\alpha=0=z$ at the 
normal phase $\lambda<\lambda_c$ (no photons and no excited atoms) but splits into two differentiated packets  at the superradiant phase $\lambda>\lambda_c$. In momentum 
space, the Husimi distribution becomes more and mode delocalized with the emergence of multiple 
modulations above the critical point $\lambda_c$ (see also later on Figure \ref{husimifig2} for a contour line of 
the variational case).

\begin{figure}
\begin{center}
\includegraphics[width=6.5cm,angle=-90]{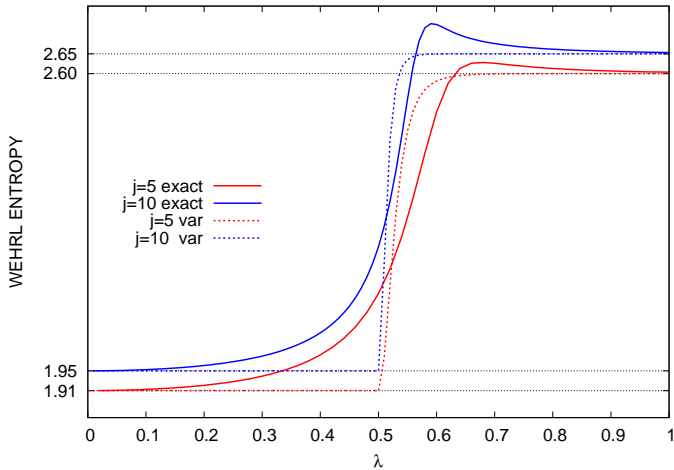}
\end{center}
\caption{(Color online) Exact (solid) and variational (dotted) Wehrl entropies  $W_j(\lambda)$ for
  $j=5$ and $j=10$ ($\omega_0=\omega=1\Rightarrow \lambda_c=0.5$) as a function of $\lambda$. Wehrl entropy $W_j(\lambda)$ grows with the number of atoms  $N=2j$ and the 
control parameter $\lambda$, attaining the limit values  $W_j(0)=1+2j/(2j+1)\stackrel{j\to\infty}{\longrightarrow} 2$ and $W_\infty(\infty)=2+\ln(2)$ in the thermodynamic limit. }
\label{wehrlfig}
\end{figure}

This delocalization of the exact Husimi distribution is captured by the  Wehrl
entropy $W_j(\lambda)$ as a function of $\lambda$ for different values of $j$. The computed results
are given in Fig. \ref{wehrlfig}, where we present  $W_j(\lambda)$ for
$j=5$ and  $j=10$ (solid lines) and for $\omega=\omega_0=1$ (for which $\lambda_c=0.5$), together with the variational results (see later).  
The Wehrl entropy tends to  $2$ (for high $j$) in the normal phase, and to  
$2+\ln 2$ in the superradiant phase, with an abrupt change (more abrupt as $j$ increases) around the
critical point.

 The exact values of $W_j(\lambda)$ for $\lambda\ll\lambda_c$ and $\lambda\gg\lambda_c$ are nicely reproduced by the following 
trial states expressed in terms of ``parity-symmetry-adapted'' CSs introduced by
Casta\~nos et al. \cite{casta1,casta2}, which turn out to be an excellent approximation to the exact quantum solution of
 the ground  (+) and first excited (--) states of the Dicke model.

Using the direct product $|\alpha,z\rangle\equiv |\alpha\rangle\otimes|z\rangle$ as 
a ground-state ansatz, one can easily compute the mean energy 
\begin{equation} \begin{array}{l}{\cal H}(\alpha,z)=\langle \alpha, z|H|\alpha,z\rangle\\
                  =\omega|\alpha|^2+j\omega_0\frac{|z|^2-1}{|z|^2+1}+{\lambda}{\sqrt{2j}}(\alpha+\bar\alpha)\frac{\bar z+z}{|z|^2+1},\\
                 \end{array}\label{ensurf}
\end{equation}
which defines a four-dimensional ``energy surface''. Minimizing with respect to these 
four coordinates gives the equilibrium points (see \cite{casta1,casta2}):
\begin{eqnarray}
\alpha_e&=&\left\{\begin{array}{ll} 0, & \mathrm{if}\, \lambda<\lambda_c\\
-\sqrt{2j}\sqrt{\frac{\omega_0}{\omega}}\frac{\lambda}{\lambda_c}\sqrt{1-\left(\frac{\lambda}{\lambda_c}\right)^{-4}}, &
\mathrm{if}\, \lambda\geq\lambda_c\end{array}\right.\nonumber\\
z_e&=&\left\{\begin{array}{ll} 0, & \mathrm{if}\, \lambda<\lambda_c\\
\sqrt{\frac{\frac{\lambda}{\lambda_c}-\left(\frac{\lambda}{\lambda_c}\right)^{-1}}{\frac{\lambda}{\lambda_c}+
\left(\frac{\lambda}{\lambda_c}\right)^{-1}}},
&
\mathrm{if}\, \lambda\geq\lambda_c\end{array}\right.\label{critpoints}
\end{eqnarray}
Note that $\alpha_e$ and $z_e$ are real and non-zero above the critical point $\lambda_c$ (i.e., in the superradiant phase).

Although the direct product  $|\alpha,z\rangle$ gives a good variational approximation to the ground state mean energy in the 
thermodynamic limit $j\to\infty$, 
it does not capture the correct behavior for other ground state 
properties sensitive to the parity symmetry $\hat\Pi$ of the Hamiltonian
(\ref{qpt01}) like, for instance, uncertainty and entropy measures \cite{epl2012,renyipra}.
This is why parity-symmetry-adapted coherent states are introduced. Indeed, a far better variational description 
of the ground (resp. first-excited) state is given in terms of the even-(resp. odd)-parity coherent states \cite{casta1,casta2}
\begin{equation}
|\psi_\pm\rangle=|\alpha,z,\pm\rangle=\frac{|\alpha\rangle\otimes|z\rangle\pm|-\alpha\rangle\otimes|-z\rangle}{{\mathcal N}_\pm(\alpha,z)},\label{sacs}
\end{equation}
obtained by applying projectors of even and odd parity $\hat{\cal P}_\pm=(1\pm \hat\Pi)$ to the direct product $|\alpha\rangle\otimes|z\rangle$. 
Here 
\begin{equation}
{\mathcal N}_\pm(\alpha,z)=\sqrt{2}\left(1\pm e^{-2|\alpha|^2}\left(\frac{1-|z|^2}{1+|z|^2}\right)^{2j}\right)^{1/2}
\end{equation}
is a normalization 
factor. These even and odd coherent states are ``Schr\"odinger cat states'' in the sense that they are a quantum 
superposition of quasi-classical, macroscopically distinguishable states. 
The new energy surface $
 {\cal H}_\pm(\alpha,z)=\langle \alpha, z,\pm|H|\alpha, z,\pm\rangle$ (see \cite{casta1,casta2} for an explicit expression of it) 
is  more involved than ${\cal H}(\alpha,z)$ in \eqref{ensurf} and makes much more difficult to obtain the new critical points 
$\alpha_e^{(\pm)},z_e^{(\pm)}$ minimizing the corresponding energy surface. The reader is addressed to Ref. \cite{castaCEWQO-12} 
in this volume for a numerical 
computation of the new critical points. It should be emphasized  that the equilibrium points given in the expression \eqref{critpoints} 
are correct only in the thermodynamic limit $j\to\infty$ or far from $\lambda=\lambda_c$ for finite $j$. 
Otherwise the minimization of ${\cal H}_\pm(\alpha,z)$  should be done (see \cite{casta1,casta2,castaCEWQO-12} 
for more details). In this paper, instead of carrying out a numerical computation of 
$\alpha_e^{(\pm)},z_e^{(\pm)}$ for different values of $j$ and $\lambda$, we shall use the approximation  
$\alpha_e^{(\pm)}\approx \alpha_e, z_e^{(\pm)}\approx z_e$, which turns out to be quite good 
except in a close neighborhood around $\lambda_c$, which diminishes as the
number of particles $N=2j$ increases (see Refs.
\cite{casta2,castaCEWQO-12}). 
With this approximation, we expect a rather good agreement between our numerical
and variational results except perhaps in a close vicinity of $\lambda_c$ (indeed, see Figure \ref{wehrlfig}).

Taking into account the coherent state overlaps
\begin{eqnarray}
 \langle\alpha|\pm\alpha_e\rangle&=&e^{-\frac{1}{2}|\alpha|-\frac{1}{2}\alpha_e^2\pm\bar\alpha\alpha_e},\nonumber\\ 
 \langle z|\pm z_e\rangle&=&\frac{(1\pm\bar zz_e)^{2j}}{(1+|z|^2)^j(1+z_e^2)^j},
\end{eqnarray}
the Husimi distribution for the variational states $|\alpha_e,z_e,\pm\rangle$ can be simply written as:
\begin{equation}
 {\Psi}_\pm(\alpha,z)=\frac{|\langle\alpha,z|\alpha_e,z_e\rangle\pm\langle\alpha,z|-
\alpha_e,-z_e\rangle|^2}{{\mathcal N}_\pm^2(\alpha_e,z_e)}.\label{Husimij}
\end{equation}
From now on we shall restrict ourselves to the even case and simply denote by 
${\Psi}={\Psi}_+$ the Husimi distribution of the variational ground state. Figure \ref{husimifig2} shows a contour line of the variational Husimi distribution. Note that, 
in position space, it reproduces the packet splitting across the critical point depicted in Figure \ref{husimifig}, 
with two differentiated packets located around the equilibrium points ($\alpha_e,z_e)$ 
and its antipode $(-\alpha_e,-z_e)$ in the superradiant phase. In momentum space, it exhibits a delocalization and `modulation'  for 
increasing values of $\lambda$.
We can easily compute the Wehrl entropy of \eqref{Husimij}, which gives 
\begin{equation}
W_j(\lambda)=\left\{\begin{array}{ll} 1+\frac{2j}{2j+1}, & \mathrm{if}\, \lambda< \lambda_c\\
1+\frac{2j}{2j+1}+\ln 2, &
\mathrm{if}\, \lambda\gg \lambda_c.\end{array}\right.
\end{equation}
denoting an entropy excess of $\ln(2)$ in the superradiant phase. In the normal phase we have exactly $W_j(\lambda)=1+{2j}/({2j+1})$, 
as corresponds to a coherent state according 
to the (still unproved) Lieb's conjecture. Indeed, as conjectured by Wehrl \cite{Wehrl} 
and proved by Lieb \cite{Lieb}, any Glauber coherent state $|\alpha\rangle$ has a minimum Wehrl entropy of 1. 
In the same paper by Lieb \cite{Lieb}, it was also conjectured that the extension of
Wehrl's definition of entropy for coherent spin-$j$ states $|z\rangle$ will yield a minimum entropy of $2j/(2j+1)$. For the joined 
system of radiation field plus atoms we would have $W_j(\lambda)=1+2j/(2j+1)$ in the normal phase ($\lambda<\lambda_c$), and therefore, 
$W_j\to 2$ in the thermodynamic limit $j\to\infty$, in agreement with our result. 

\begin{figure}
\begin{center}
\includegraphics[width=7cm]{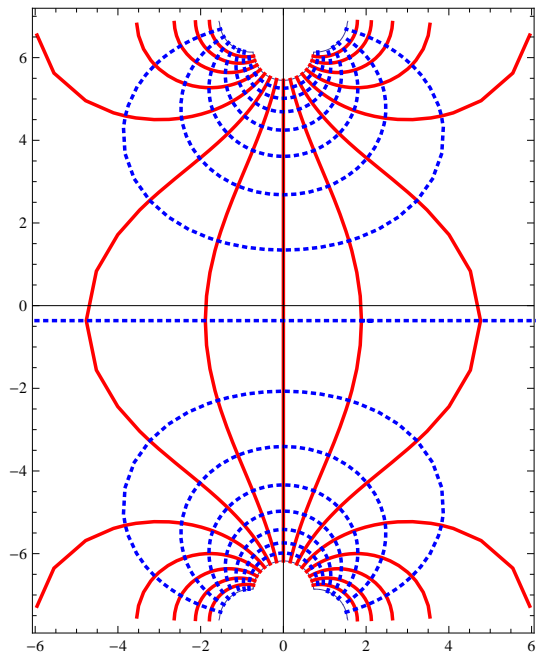}
\end{center}
\caption{(Color online) Surface of zeros $\alpha=f_j^{(l)}(z,\lambda)$ of the variational Husimi distribution $\Psi_+(\alpha,z)$ 
for $\lambda=1, j=10$ and $l=0$  ($\lambda_c=0.5$) seen as a conformal 
mapping of a regular grid in the $z$-plane.}
\label{confmap}
\end{figure}

To finish, we would like to comment on the zeros of the Husimi distribution as a fingerprint for the QPT (see \cite{husidi} for more information). 
From \eqref{Husimij} we obtain
\begin{equation}
 {\Psi}(\alpha,z)=0\Rightarrow 2\alpha\alpha_e+2j\ln\frac{1+ z z_e}{1- z z_e}=i\pi (2l+1),\,l\in\mathbb Z,\label{zr}
\end{equation}
which defines  a two-dimensional surface (for each value of $l$)  in a four-dimensional manifold with parametric equations:
\begin{equation}
 \alpha=f_j^{(l)}(z,\lambda)=\frac{j}{\alpha_e}\ln\frac{1- z z_e}{1+ z z_e}+\frac{i\pi}{2\alpha_e} (2l+1). \label{zerosfj}
\end{equation}
This expression gives  in particular the ``less probable mean photon number $|\alpha|^2$ for each 
mean atom fraction $|z|^2/(1+|z|^2)$'' in phase space (remember 
comment before Eq. \eqref{probamp}). In Figure 
\ref{confmap} we represent this surface as a conformal mapping of a regular grid in the $z$-plane. That is, 
for $z=z_1+iz_2$, we represent the image of vertical lines $z_1=$constant (solid-red curves) and horizontal lines 
$z_2=$constant (dotted-blue curves).
We see from \eqref{zr} that, in the normal phase ($\alpha_e=0=z_e$) the Husimi distribution $\Psi(\alpha,z)$ has no zeros. In the 
superradiant phase ($\lambda>\lambda_c$) there are more and more zeros as $j$ and $\lambda$ increase. To study the high $j$ limit, we 
can redefine $\beta\equiv\sqrt{2j}\, z$, which simplifies the expression of:
\begin{equation}
 2j\ln\frac{1+ z z_e}{1- z z_e}\simeq 2\beta\beta_e,\;\;\hbox{for}\; j\gg 1,
\end{equation}
where we have made use of the definition of the Euler number at some stage. Therefore, the equation \eqref{zerosfj} becomes:
\begin{equation}
 \alpha_1=-\frac{\beta_e}{\alpha_e}\beta_1,\;
 \alpha_2=-\frac{\beta_e}{\alpha_e}\beta_2+\frac{\pi}{2\alpha_e}(2l+1).\label{zeros2}
\end{equation}
for $\alpha=\alpha_1+i\alpha_2$ and $\beta=\beta_1+i\beta_2$. Therefore, in the high $j$ limit, and in the 
superradiant phase ($\lambda>\lambda_c$), the zeros are localized along straight lines (``dark fringes'') in the 
$\alpha_1\beta_1$ (position) and $\alpha_2\beta_2$ (momentum) planes. In the momentum plane, the number of dark 
fringes  grows with $\lambda$ and $j$. In the 
thermodynamic limit $j\to\infty$, zeros densely fill the momentum plane 
$\alpha_2\beta_2$ (see \cite{husidi} for a graphical representation of zeros in the high $j$ limit). 

\section{Conclusions}\label{sec3}

We have found that Wehrl entropy of the Husimi distribution provides a 
sharp indicator of a quantum phase transition in the Dicke model even for finite $j$. This uncertainty measure detects a delocalization of the 
Husimi distribution across the critical point $\lambda_c$ and we have employed it, together with three-dimensional plots and contour lines of the 
Husimi distribution,  to quantify and visualize the phase-space spreading of
the ground state. 

Calculations have been done numerically and through a variational approximation.  The variational approach, in terms of symmetry-adapted 
coherent states, complements and 
enriches the analysis  providing explicit analytical
expressions for the Husimi distribution and Wehrl entropies which remarkably coincide with the
numerical results, especially in the thermodynamic limit and far from $\lambda=\lambda_c$, where the approximate equilibrium points \eqref{critpoints} fail. A more 
accurate calculation could be perhaps done by using the `true' equilibrium points of Ref. \cite{castaCEWQO-12}, although we think our variational approach still 
captures the qualitative behavior near $\lambda_c$ and the quantitative  exact values far from $\lambda_c$ (see again Figure \ref{wehrlfig} in this respect).

In the superradiant phase, 
Wehrl entropy undergoes an 
entropy excess  of $\ln(2)$. This fact implies that the Husimi distribution splits 
up into two identical subpackets with negligible overlap in passing from normal to superradiant phase. In general, for 
$s$ identical subpackets with negligible
overlap, one would expect an entropy excess of $\ln(s)$.

The QPT fingerprints in the Dicke model have also been tracked by exploring
the distribution of zeros of the Husimi density within the
  analytical variational approximation. Now, we have corroborated  that the
  zeros of the Husimi distribution evidence the QPT
  without the Holstein-Primakoff approximation, founding again
  that there are no zeros in  the normal phase and  a larger number of zeros as 
 $j$ and $\lambda$ increase in the superradiant phase. This interesting result
  supports the asseveration that the emergence of zeros of the  Husimi
  distribution can  be   an
  indicator of QPTs \cite{husidi,husivi}. 

\section*{Acknowledgments}

This work was supported by the Projects:   FIS2011-24149 and FIS2011-29813-C02-01 (Spanish MICINN),  
FQM-165/0207 and FQM219 (Junta de Andaluc\'\i a)  and 20F12.41 (CEI BioTic
UGR).

\end{document}